\documentclass[a4paper,12pt]{article}
\usepackage[utf8x]{inputenc} 
\usepackage{jheppub}
\usepackage{amsmath}
\usepackage{slashed}
\usepackage{axodraw}
\usepackage[T1]{fontenc}
\usepackage{comment}
\usepackage{hyperref}
\usepackage{braket}
\usepackage{graphicx}
\usepackage{subfigure}

\title{ \boldmath Electroweak Baryogenesis and Dark Matter via a Pseudoscalar vs. Scalar}

\author[]{Parsa Hossein Ghorbani}

\affiliation[]{Institute for Research in Fundamental Sciences (IPM),\\ 
 School of Particles and Accelerators, P.O. Box 19395-5531, Tehran, Iran}

\emailAdd{pghorbani@ipm.ir}

\date{}
\abstract{We study the electroweak baryogenesis in a fermionic dark matter scenario with a (pseudo)scalar 
being the mediator in the Higgs portal. It is discussed that the electroweak phase 
transition turns to be first-order after taking into account the role of the (pseudo)scalar in the thermal
effective potential in our extended standard model. Imposing the relic density constraint from the WMAP/Planck 
and the bounds from the direct detection experiments XENON100/LUX, we show that the dark matter scenario with a scalar 
mediator is hardly capable of explaining the baryogenesis while the same model with a pseudoscalar mediator 
is able to explain the baryon asymmetry. For the latter, we constrain more the model with {\it Fermi}-LAT upper 
limit on dark matter annihilation into $b\bar b$ and $\tau^+\tau^-$. The allowed dark matter mass that leads to 
correct relic abundance, renders the 
electroweak phase transition strongly first-order, and respects the {\it Fermi}-LAT limit,
will be in the range $110-320$ GeV. The exotic and invisible Higgs decay bounds and the mono-jet
search limit at the LHC do not affect the viable space of parameters.}


\begin{document}
\maketitle

\section{Introduction}
The electroweak symmetry breaking which we believe is a cornerstone in the Higgs physics could be restored at 
high temperature i.e. in the early universe \cite{Weinberg:1974hy,Dolan:1973qd}. The spontaneous symmetry 
breaking occurs by having the universe expanded and cooled and going from 
the symmetric phase to the broken phase through the {\it electroweak phase transition} (EWPT). 

An open problem in physics is the mysterious matter-antimatter asymmetry in the universe. The standard model (SM) together
with the EWPT seem to be able to explain the exceed of the matter abundance against the antimatter 
we observe today. This is 
called the {\it electroweak baryogenesis} (EWBG). The necessary conditions that lead to EWBG is known as the three criteria of 
Sakharov \cite{Sakharov:1967dj}: Baryon number violation, C and CP violation and the thermal non-equilibrium.
The standard model possesses all these conditions however 
at the temperature of the electroweak symmetry breaking, $T_{\text{EW}}\sim 100$ GeV, the electroweak phase transition 
must be of first-order type in order to support the EWBG. This can be translated into {\it washout criterion}, 
$v(T_c)/T_c>1$ which provides the appropriate sphaleron rate for the baryogenesis. In the standard model framework
the washout condition impose an upper limit on the Higgs mass: $m_H<48$ GeV \cite{Shaposhnikov:1986jp} which is in contrast
with the Higgs discovery ($m_H\sim 125$ GeV) at the LHC in 2012 \cite{Aad:2012tfa,Chatrchyan:2012xdj}. Therefore the SM 
alone is not enough to embed the EWBG and an extension to SM looks necessary if the electroweak baryogenesis is the 
right mechanism to produce the baryon asymmetry.

On the other hand, almost $25\%$ of the universe is made of what we call it {\it dark matter} (DM). It 
does not have strong interaction with the ordinary matter and its existence can be traced only through its
gravitational attraction on matter \cite{Adam:2015rua,Hinshaw:2012aka}. Again the SM is not capable of explaining 
this extra gravitational force in the galaxies and clusters. Inevitably one should search for an answer in a 
theory beyond the standard model. One of the successful DM scenarios is the {\it weakly interacting massive particle}
(WIMP) which demands additional fundamental fields (particles) respect to the SM. The new particle(s) stay
in thermal equilibrium with other SM particles in the early universe, 
but when the universe cools down while expanding, it freezes-out from the hot plasma of particles. The 
self-annihilation cross section of the WIMP must be of order $\braket{\sigma v}=3\times 10^{-26}~cm^3 ~s^{-1}$   
to have the correct amount of DM abundance to be $\Omega h^2\sim 0.11$ measured by Planck/WMAP. 

In this work we try to address both the electroweak baryogenesis and the dark matter issues in one single theory. 
We extend the standard model by adding an extra Dirac fermion playing the role of the dark matter candidate and a 
(pseudo)scalar 
mediating between the dark sector and the SM sector \cite{Ghorbani:2014qpa}.
\footnote{See \cite{Ghorbani:2017qwf} for the analysis of the renormalization group equation for the same model with 
a pseudoscalar mediator.}
The (pseudo)scalar and the fermionic 
dark matter modify the thermal effective potential; hence affect the critical temperature for the electroweak
phase transition. We analytically provide the critical temperature and the global minimum at the critical temperature
to give the washout criterion. Many works have provided the critical temperature from the free energy by numerical 
computation. Some examples of the works done considering both the dark matter and the baryogenesis are 
\cite{Gu:2010ft,Cline:2012hg,Fairbairn:2013xaa,Lewicki:2016efe,
Jiang:2015cwa,Chowdhury:2011ga,Menon:2004wv,Barger:2008jx,Li:2014wia,Addazi:2017gpt}. 
Here we provide an analytical expression for the critical temperature
which makes it easier to impose the washout condition on the dark matter model. 

If the SM-DM mediator is a pseudoscalar the dark matter cross section off nucleon is negligible while if the 
mediator is a scalar then the direct detection experiments e.g. XENON100/LUX put a strong constraint on the space parameter.  
Having emphasized this point, we show in this work that the dark matter model with a pseudoscalar mediator is more successful 
in explaining the electroweak baryogenesis respect to when we use the scalar mediator. 

This paper is organized as the following: in the next section we introduce the model by extending the SM. 
In section \ref{ewpt} the details of the electroweak phase transition and the critical temperature are given. In
section \ref{dm} the dark matter candidate is introduced. In
the next section, the numerical results that shows the consistency of the model with relic
density and the baryon asymmetry is presented. The appendix \ref{1loop} provides 
some analytical details on the effective potential.

\section{The Model}\label{model}

We extend the standard model
by adding two new fields as follows: a Dirac fermion denoted here by $\psi$, which plays the role 
of the dark matter candidate, and a (pseudo)scalar denoted by $s$ which 
mixes with the Higgs field, $H$, and interacts with the dark matter as well through a Yukawa term. 
The Lagrangian can be written in its parts as,
\begin{equation}
{\cal L}={\cal L}_{\text{SM}}(H)+{\cal L}_{\text{dark}}(\psi)+{\cal L}_{s}(s)+{\cal L}_{\text{int}}\left(s,\psi,H\right)\,,
\end{equation}
 where $\mathcal{L}_{\text{SM}}$ stands for the SM Lagrangian, ${\cal L}_{\text{dark}}$ 
 for the fermionic dark matter Lagrangian, 
 
\begin{equation}\label{darkl}
{\cal L}_{\text{dark}}\left(s,\psi\right)=\bar{\psi}\left(i\gamma_{\mu}\partial^{\mu}-m_{d}\right)\psi \, ,
\end{equation}  
${\cal L}_{s}$  for the (pseudo)scalar Lagrangian, 

\begin{equation}
{\cal L}_{s}=\cfrac{1}{2}\left(\partial_{\mu}s\right)^{2}-\frac{1}{2}\mu_{s}^{2}s^{2}-\frac{1}{4}\lambda_{s}s^{4}\, ,
\end{equation}
and ${\cal L}_{\text{int}}$ is the (pseudo)scalar interaction with the dark and the SM sectors. When $s$ is a
pseudoscalar then, 

\begin{equation}\label{intl1}
{\cal L}_{\text{int}}\left(s,\psi,H\right)=g_{d}s\bar{\psi}\gamma^{5}\psi+\frac{1}{2}\lambda s^{2}H^{\dagger}H \, ,
\end{equation}
and when $s$ represents the scalar, 

\begin{equation}\label{intl2}
{\cal L}_{\text{int}}\left(s,\psi,H\right)=g_{d}s\bar{\psi}\psi+\frac{1}{2}\lambda s^{2}H^{\dagger}H \, .
\end{equation}
In the next sections we investigate both the scalar and the pseudoscalar cases and use 
the same notation for the coupling $g_d$. 

Note that the interaction Lagrangian does not include the 
odd-terms in the (pseudo)scalar-Higgs interaction terms. In order to stay in a more restricted theory 
with as small parameter space as possible, despite many
authors we consider a less general Lagrangian for our model. The Higgs potential in the SM sector reads,
\begin{equation}
V_{H}=-\mu_{h}^{2}H^{\dagger}H-\lambda_{h}\left(H^{\dagger}H\right)^{2}\,.
\end{equation}
Both the Higgs and the (pseudo)scalar take
non-zero vacuum expectation values at the low temperature. In section \ref{ewpt} 
to give the thermal effective potential as a function of the condensate $\braket{h}$
we begin with the tree-level potential which is given by the substitution 
$s \equiv \braket{s}$ and $H=\left(0\,\,\,h\equiv \braket{h} \right)^{\dagger}$
So the tree-level potential reads, 
\begin{equation}\label{v0}
V_{0}(h,s)=-\frac{1}{2}\mu_{h}^{2}h^{2}-\frac{1}{2}\mu_{s}^{2}s^{2}+\frac{1}{4}\lambda_{h}h^{4}
+\frac{1}{4}\lambda_{s}s^{4}+\frac{1}{2}\lambda s^{2}h^{2}.
\end{equation}
Note that we have gauged away three degrees of freedom of the Higgs doublet.

\section{First-Order Phase Transition}\label{ewpt}

In this section we provide the ``washout criterion''  and other necessary conditions 
in terms of the parameters used in the 
model introduced above 
to support the first-order electroweak phase transition. The washout criterion or $v(T_c)/T_c>1$ 
which provides the appropriate sphaleron rate for the phase transition to be first-order is obtained 
from the effective potential of the theory. In addition to the tree-level barrier we also consider the 
one-loop barrier. The total thermal effective potential is therefore, 

\begin{equation}
 V_{\text{eff}}=V_0(h,s)+V_{\text{1-loop}}(h,s;0)+V_{\text{1-loop}}(h,s;T)\,,
\end{equation}
where $V_0$ is the tree-level potential in eq. (\ref{v0}), $V^{\text{1-loop}}(h,s;0)$ is the Coleman-Weinberg 
one-loop correction 
at zero temperature \cite{Coleman:1973jx} and $V^{\text{1-loop}}(h,s;T)$ is the 
one-loop thermal correction \cite{Dolan:1973qd}. In the high-temperature approximation 
when $m_i^2/T^2 \ll 1$ for all $i$ with $m_i$ being the mass of the particle $i$ in the model, 
the one-loop effective potential takes the following form,

\begin{equation}\label{highT}
V_{\text{1-loop}}^{\text{high-T}}\left(h,s;T\right) \approx \left(\frac{1}{2}\kappa_{h}h^{2}
+\frac{1}{2}\kappa_{s}s^{2}\right)T^{2}\,,
\end{equation}
with 
\begin{equation}\label{kaph}
\kappa_{h}=\frac{1}{48}\left(9g^{2}+3g'^{2}+12g_{t}^{2}+24\lambda_{h}+4\lambda\right)\,,
\end{equation}
\begin{equation}\label{kaps}
\kappa_{s}=\frac{1}{12}\left(4\lambda+3\lambda_{s} \pm 2g_{d}^{2}\right)\,, 
\end{equation}
where in eq. (\ref{kaps}) the minus sign stands for the pseudoscalar case and the plus sign is for the scalar case.
See appendix \ref{1loop} for more details. 

Note that we have dropped the Colman-Wienberg zero-temperature correction since at high temperature approximation 
(at temperature of the electroweak phase transition) only the thermal corrections are dominant. Moreover, including
the zero-temperature contribution will only complicate the analytic computations.

In eq. (\ref{kaph}) the parameters $g$ and $g'$ are respectively the $SU(2)_L$ and $U(1)_Y$ standard model couplings and $g_t$
is the top quark coupling. 
We have contributed only the heavier particles that couple stronger to the Higgs, i.e. the top quark, $t$, 
the gauge bosons, $W^{\pm}$ and $Z$, and the Higgs, $h$. 
We have ignored the lighter quarks, gluons and the leptons.

In order to have a first-order phase transition the effective potential
must have two degenerate minima at the critical temperature. The minima
are located at 
\begin{equation}
\frac{\partial V_{\text{eff}}}{\partial s}|_{w\left(T\right)=<s>}=0\,\,\,\,\,\,\,\,\,\,\,\,\,\,\,\,\,\,\,\,\,\,\
\frac{\partial V_{\text{eff}}}{\partial h}|_{v\left(T\right)=<h>}=0\,,
\end{equation}
where leads to 
\begin{equation}
v_{\text{sym}}=0\,\,\,\,\,\,\,\,\,\,\text{or}\,\,\,\,\,\,\,\,\,\,\,v_{\text{brk}}^{2}\left(T\right)
=\frac{\mu_{h}^{2}-\kappa_{h}T^{2}-\lambda w^{2}\left(T\right)}{\lambda_{h}}\label{vev1}\,,
\end{equation}
\begin{equation}
v_{\text{brk}}^{2}\left(T\right)=\frac{\mu_{s}^{2}-\kappa_{s}T^{2}-\lambda_{s}w^{2}\left(T\right)}{\lambda}\label{vev2}\,.
\end{equation}
The stability conditions are obtained by setting the positivity of
the second derivatives of the potential after the symmetry breaking:
\begin{equation}
\lambda_{s}>0,\,\,\,\,\,\lambda_{h}>0,\,\,\,\,\,4\lambda_{s}\lambda_{h}>\lambda^{2}\label{stab}\,.
\end{equation}
Eqs. (\ref{vev1}) and (\ref{vev2}) leads to 
\begin{equation}
w_{\text{brk}}^{2}\left(T\right)=\alpha+\beta T^{2}\,,
\end{equation}
where 
\begin{equation}
\alpha=\frac{\lambda\mu_{h}^{2}-\lambda_{h}\mu_{s}^{2}}{\lambda^{2}-\lambda_{h}\lambda_{s}}\,,
\end{equation}
\begin{equation}
\beta=\frac{\lambda\kappa_{h}-\lambda_{h}\kappa_{s}}{\lambda^{2}-\lambda_{h}\lambda_{s}}\,.
\end{equation}
Using eq. (\ref{vev1}) one gets the temperature-dependent Higgs vacuum expectation value in the broken phase, 
\begin{equation}\label{vc}
v_{\text{brk}}^{2}\left(T\right)=\left(\frac{\mu_{h}^{2}-\lambda\alpha}{\lambda_{h}}\right)-\left(\frac{\kappa_{h}
+\lambda\beta}{\lambda_{h}}\right)T^{2}\,.
\end{equation}
The broken phase can exist up to a maximal temperature,
\begin{equation}
 T_{\text{max}}=\sqrt{\frac{\mu_{h}^{2}-\lambda\alpha}{\kappa_{h}+\lambda\beta}},\,\,\,\,\,\,\mu_{h}^{2}>\lambda\alpha\,.
\end{equation}

At the critical temperature the free energy has two degenerate minima
one at \\ $\left(v_{\text{sym}}=0,w_{\text{sym}}=0\right)$ and the other
at $\left(v_{\text{brk}},w_{\text{brk}}\right)$, 
\begin{equation}\label{freee}
V_{\text{eff}}^{\text{sym}}\left(T_{c}\right)=V_{\text{eff}}^{\text{brk}}\left(T_{c}\right)=0\,,
\end{equation}
which solves the free energy eq. (\ref{freee}) at the critical temperature as, 
\begin{equation}\label{Tc}
T_c^{\pm}= \sqrt{x\pm y^{1/2}/z}\,,
\end{equation}
with 
\begin{equation}\label{xyz}
 \begin{split}
&x= {\kappa_h} {\lambda_h} (\lambda  {\mu^2_s}-{\lambda_s} {\mu^2_h})
+{\kappa_s} {\lambda_h} (\lambda  {\mu^2_h}-{\lambda_h} {\mu^2_s})\,,\\
&y= {\kappa_h}^2 \lambda_h^2 \left(-8 \lambda ^3 {\mu^2_h} {\mu^2_s}
+ \lambda ^2    \left(5 {\lambda_h} {\mu_s}^4+4 {\lambda_s} {\mu_h}^4\right)-{\lambda_h}^2    {\lambda_s} {\mu_s}^4\right)\\
& +2 {\kappa_h} {\kappa_s} {\lambda_h^3} \left(7 \lambda ^2    {\mu^2_h} {\mu^2_s}
-4 \lambda  \left({\lambda_h} {\mu_s}^4+{\lambda_s} {\mu_    h}^4\right)+{\lambda_h} {\lambda_s} {\mu^2_h} {\mu^2_s}\right)\\
&+{\kappa_s}^2    {\lambda_h^3} \left(\lambda ^2 {\mu_h}^4-8 \lambda  {\lambda_h} {\mu^2_h} {\mu^2_    s}
+{\lambda_h} \left(4 {\lambda_h} {\mu_s}^4+3 {\lambda_s} {\mu_    h}^4\right)\right)\,,\\
&z={\kappa_h}^2 \left(4 \lambda ^2-{\lambda_h} {\lambda_s}\right)
-6    {\kappa_h} {\kappa_s} \lambda  {\lambda_h}+3 {\kappa_s}^2 {\lambda_h}^2\,.
 \end{split}
\end{equation}
The strong electroweak phase transition occurs only if 
\begin{equation}\label{wash}
v\left(T_{c}\right)/T_{c}>1\,,
\end{equation}
which is a strong constraint on the parameter space of our dark matter model. 
Note that for each set of the couplings in eq. (\ref{xyz}) 
there are two critical
temperatures,  $T_c^{\pm}$, if $x\pm y^{1/2}/z>0$. 
We will see in section \ref{dm} that for both solutions there exist a viable parameter space. 

\section{Fermionic Dark Matter}\label{dm}
We are assuming that the first-order phase transition is occurring
in a temperature higher than the dark matter freeze-out temperature, i.e. 
$T_{c}>T_{F}$. After the symmetry breaking both Higgs particle and
the scalar field undergo non-zero {\it vev}:
\begin{equation}
h\rightarrow v_{h}+h\label{shifth}\,,
\end{equation}
\begin{equation}
s\rightarrow v_{s}+s\label{shifts}\,.
\end{equation}
This will cause to a three potential different from eq. (\ref{v0}).
One can redefine the scalar fields $h$ and $s$ in order to diagonalize
the mass matrix as follows,

\begin{equation}
\begin{split}
 h\rightarrow h\,\cos\theta+s\,\sin\theta \,, \\
s\rightarrow-h\,\sin\theta+s\,\cos\theta\,.
\end{split}
\end{equation}
For simplicity we denote the new fields as $h$ and $s$ again but
one should be noted that these are the eigenstates of the diagonal
mass matrix now. Plugging eqs. (\ref{shifth}) and (\ref{shifts})
in eq. (\ref{v0}) and imposing the minimization condition we get
\begin{equation}
\mu_{s}^{2}=-\lambda_{s}v_{s}^{2}-\lambda v_{h}^{2}\label{mus}\,,
\end{equation}
\begin{equation}
\mu_{h}^{2}=-\lambda_{h}v_{h}^{2}-\lambda v_{s}^{2} \label{muh}\,.
\end{equation}
The second derivative of the potential in eq. (\ref{v0}) after shifting
the origin to the non-zero vacuum are,

\begin{equation}
 \tilde{m}_{s}^{2}=2\lambda_{s}v_{s}^{2},\,\,\,\,\,\,\,\,\,\,\,\,\,\,\,\,\,\,\,\,\tilde{m}_{h}^{2}
=2\lambda_{h}v_{h}^{2},\,\,\,\,\,\,\,\,\,\,\,\,\,\,\,\,\,\,\,\tilde{m}_{hs}^{2}=\lambda v_{h}v_{s} \,.
\end{equation}

Diagonalizing the mass matrix by a rotation by the angle $\theta$, the
couplings will be related to the masses as, 
\begin{equation}
 \begin{split}
  \lambda_{h}=\frac{m_{s}^{2}\sin^{2}\theta+m_{h}^{2}\cos^{2}\theta}{2v_{h}^{2}}\,,\\
\lambda_{s}=\frac{m_{s}^{2}\cos^{2}\theta+m_{h}^{2}\sin^{2}\theta}{2v_{s}^{2}}\,,\\
\lambda=\frac{m_{s}^{2}-m_{h}^{2}}{2v_{h}v_{s}}\sin2\theta\,,
 \end{split}
\end{equation}
where $m_{s}^{2}$ and $m_{h}^{2}$ are the diagonalized masses and
\begin{equation}
 \tan\theta=\frac{y}{1+\sqrt{1+y^{2}}},\,\,\,\,\,\,\,\,\,\,y=\frac{2\lambda v_{s}v_{h}}{\lambda_{h}v_{h}^{2}
-\lambda_{s}v_{s}^{2}}\,.
\end{equation}

The independent parameters of the model can then be chosen as $m_{h}$,
$m_{s}$, $m_{d}$, $g_d$ and $\sin \theta$. All the other parameters including
the mass terms and the couplings in the Lagrangian are
expressed in terms of these five parameters. Note that all we have said in this section 
is true for both scalar and pseudoscalar mediators. In the next section we see how the numerical 
results changes for them. 

\section{Numerical Results}\label{nr}

Our computations are two-folded. One part is the computations for the model with a pseudoscalar mediator and the other part
is the computations with a scalar mediator. We recall that the difference they make in strongly first-order 
phase transition, section \ref{ewpt}, stem from eq. (\ref{kaps}) where the plus sign stands for the scalar and the minus 
sign stands for the pseudoscalar. In dark matter side apart from the different results for the relic density for each type 
of mediator, one should note an important point; when choosing the mediator in our dark matter scenario to be a pseudoscalar
the elastic scattering cross section for DM-nucleon is velocity suppressed 
 \cite{Esch:2013rta,LopezHonorez:2012kv,Pospelov:2011yp,Ghorbani:2014qpa} 
and so the theory easily evades the direct detection bounds from
LUX/XENON100/XENON1T \cite{Akerib:2016vxi,Aprile:2016swn,Aprile:2017iyp}. 
However this is not the case for a scalar mediator. A large region of the parameter space in the 
theory is cut by the direct detection bounds imposed from XENON100/LUX. 
Therefore, for the pseudoscalar mediator we have two 
type of constraints to impose: the relic density and the first-order phase transition condition. For the scalar mediator 
the direct detection constraint must be added to the above conditions.

To compute the relic density for the dark matter one should 
solve the Boltzmann differential equation numerically. We have exploited the {\tt MicrOMEGAs4.3}
 package \cite{Belanger:2014vza}
to obtain the relic abundance. The Higgs vacuum expectation value  and the Higgs mass are known, $v_h=246$ GeV and 
$m_h=125$ GeV.  
The {\it vev}
for the pseudoscalar and for the scalar as well is chosen to be $v_s=600$ GeV
\footnote{ 
A caveat here is that additional (pseudo)scalars in the extended SM with non-zero changing {\it vev} during the electroweak
phase transition may lead to supersonic bubble wall velocity that prevent the phase transition from 
being strong enough for EWBG 
\cite{Bodeker:2009qy,Kozaczuk:2015owa}.}.
In both cases, we search for 
the viable parameter space bounded by the 
relic density value $\Omega h^2= 0.11$ and the washout condition being $v(T_c)/T_c>1$ with $v(T_c)$ and $T_c$ given 
in eqs. (\ref{Tc}) and (\ref{vc}). We scan the space of parameters in the ranges as  GeV $1<m_s<1$ TeV, 
the dark matter mass GeV $10<m_d<2$ TeV, the 
Yukawa coupling $0<g_d<3$ and the mixing angle being fixed at $\sin \theta=0.1$.

\begin{figure}
\begin{minipage}[b]{.45\textwidth}
 \includegraphics[scale=.35, angle=270]{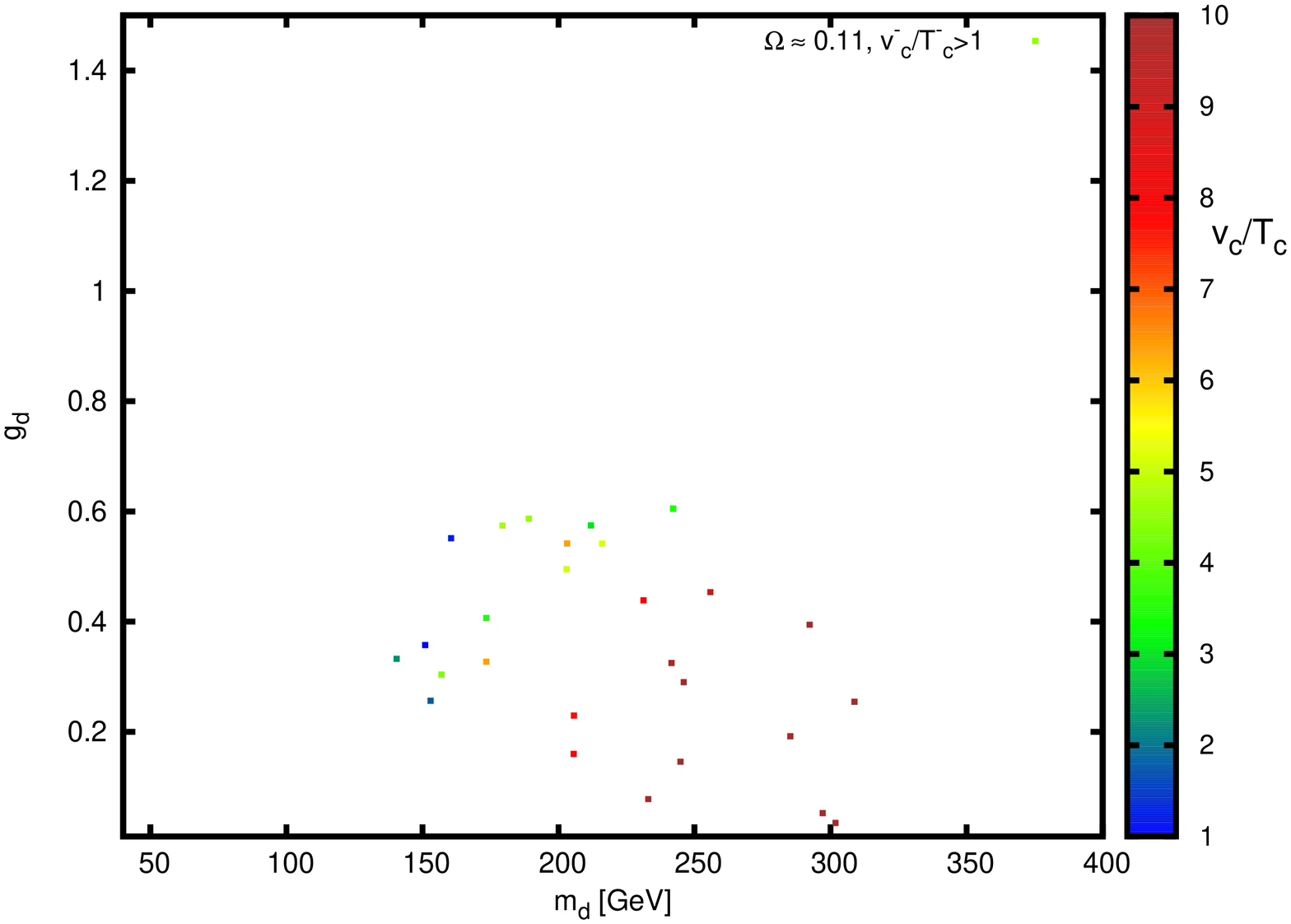}
\end{minipage}\hspace{1.5cm}
\begin{minipage}[b]{.45\textwidth}
 \includegraphics[scale=.35, angle=270]{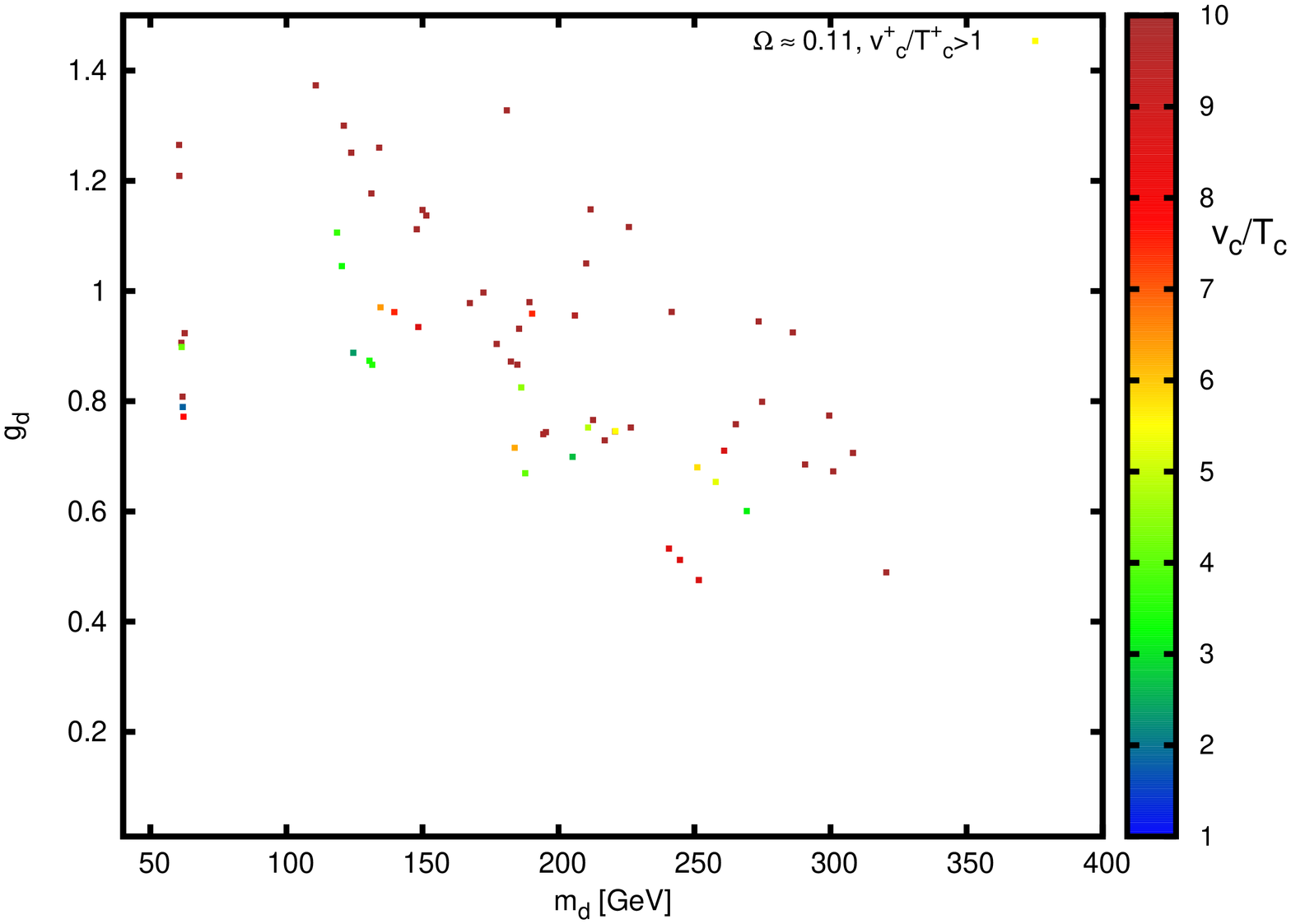}
\end{minipage}
\caption{The allowed DM mass, $m_d$, against the Yukawa coupling, $g_d$, in the fermionic dark matter scenario 
with pseudoscalar mediator, bounded from the 
relic density and first-order phase transition conditions for {\it left)} $v^-_c/T^-_c>1$ with DM mass in 
the range $140-310$ GeV {\it right)} $v^+_c/T^+_c>1$ with DM mass in the range $60-320$ GeV.
}
\label{pseudo-plot}
\end{figure}

\begin{figure}
\begin{minipage}[b]{.45\textwidth}
 \includegraphics[scale=.35, angle=270]{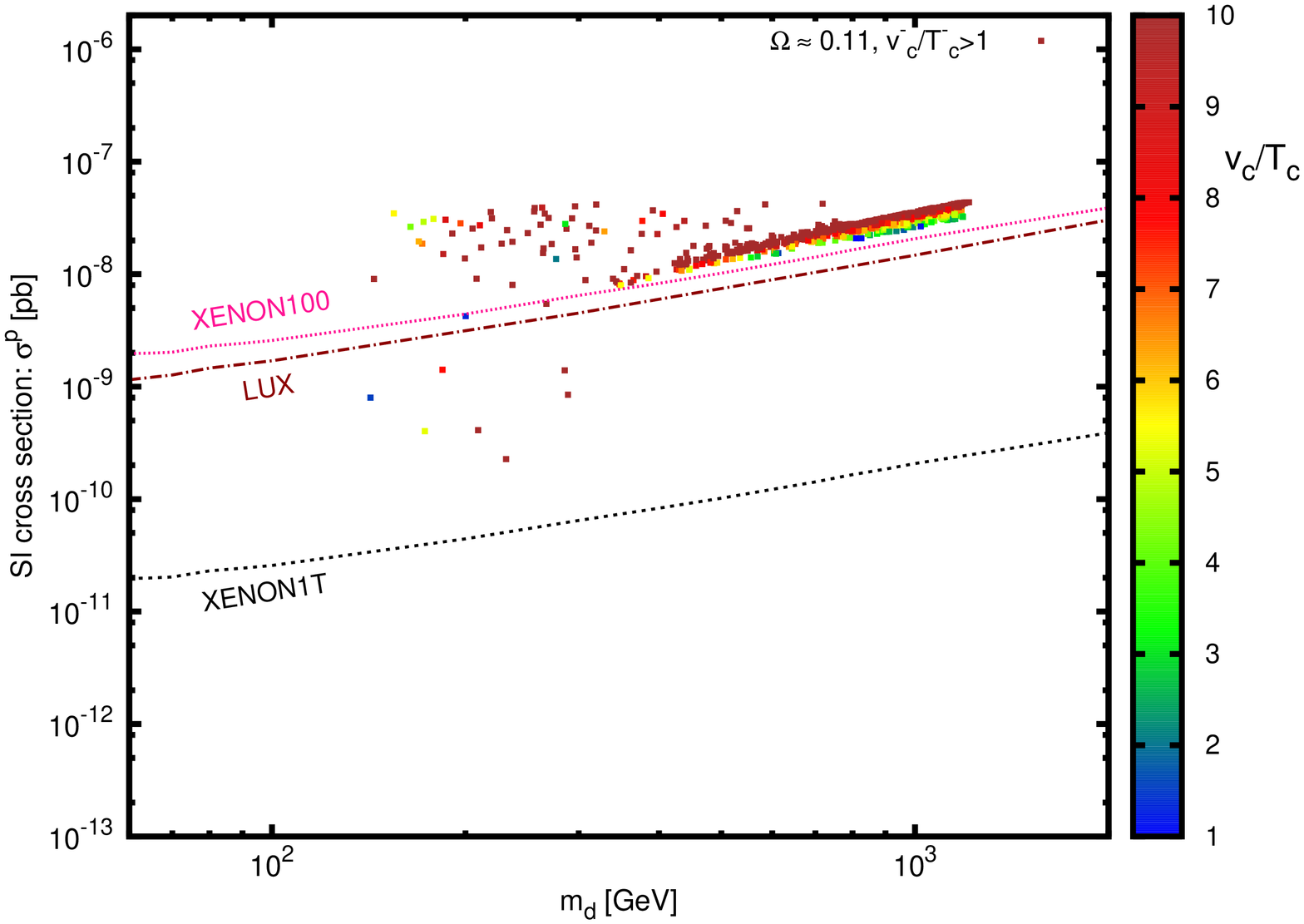}
\end{minipage}\hspace{1.5cm}
\begin{minipage}[b]{.45\textwidth}
 \includegraphics[scale=.35, angle=270]{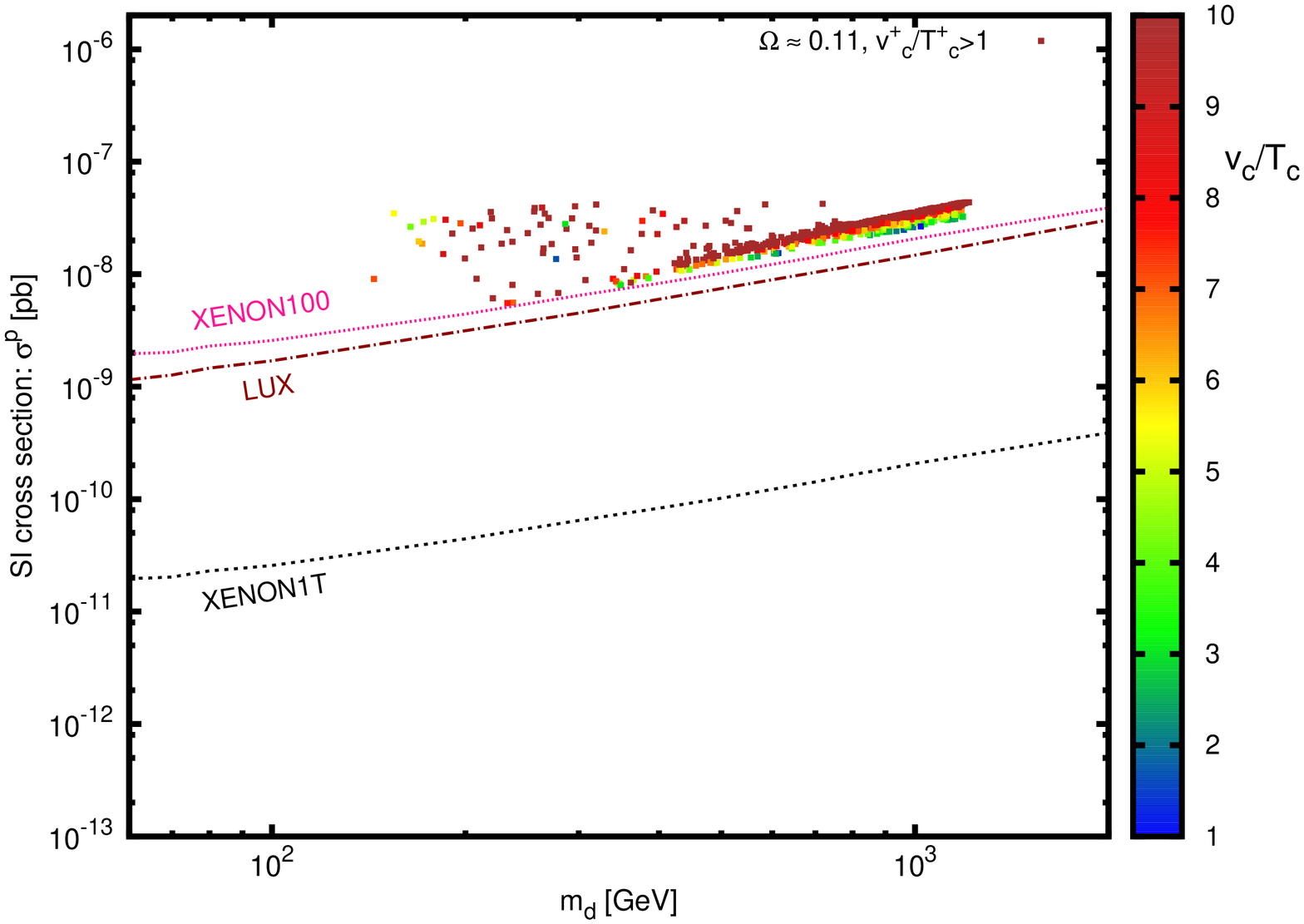}
\end{minipage}
\caption{The allowed DM mass, $m_d$, against the Yukawa coupling, $g_d$, in the fermionic dark matter scenario 
with scalar mediator, bounded from the 
relic density, first-order phase transition condition, and the direct detection bound by XENON100/XENON1T/LUX 
for {\it left)} $v^-_c/T^-_c>1$ with DM mass in the range $150-300$ GeV for XENON100/LUX and no viable DM mass for 
the future direct detection bound by XENON1T {\it right)} $v^+_c/T^+_c>1$ with no viable DM mass.}
\label{scalar-plot}
\end{figure}

As we have demonstrated in Figs. \ref{pseudo-plot} and \ref{scalar-plot} that despite the strong constraints on the 
parameter space we end up with a viable region for the theory Fig. \ref{pseudo-plot} shows the allowed DM mass
against the Yukawa coupling 
for the dark matter model with the pseudoscalar mediator and Fig. \ref{scalar-plot} represents the same quantities 
for the scalar mediator. The constraints considered in two figures are the amount of the relic abundance from WMAP/Planck, 
the washout criterion for two critical temperature found in eq. (\ref{Tc}) and the DM-nucleon elastic scattering 
cross section bound from XENON100/XENON1T/LUX. 

It is interesting to point out the difference in two figures; the fermionic 
dark matter model with a pseudoscalar mediator Fig. \ref{pseudo-plot} is more successful respect 
to the same model with a scalar mediator Fig. \ref{scalar-plot} to 
accommodate the electroweak baryogenesis and the dark matter issues. 
The allowed DM mass in Fig. \ref{pseudo-plot} is in the range GeV $60 <m_d< 320$ Gev for the $T_c^+$ solution and in the range 
GeV $140 <m_d< 310$ Gev for the $T_c^-$ solution. Therefore for both critical temperatures obtained in eq. (\ref{Tc}) 
the pseudoscalar scenario 
respect the relic density and the first-order phase transition. The allowed DM mass for the scalar scenario is 
only in the range GeV $150 <m_d< 300$ Gev for the $T_c^-$ solution which can easily be excluded 
by the future direct detection experiments such as XENON1T. This point has already been reported in \cite{Li:2014wia}. 
For $T_c^+$ solution there is no viable region in the scalar scenario.

\subsection*{Indirect Detection Bounds}
The DM annihilation into 
SM particles is very suppressed after the thermal freeze-out in the early universe. Nevertheless, today in regions
with high density of DM for instance in the Galactic Center (GC) the DM annihilation is probable. 
It is now well accepted from {\it Fermi} Large Area Telescope ({\it Fermi}-LAT) $6.5$ years data 
that the gamma rays coming from the center of the galaxies are brighter for a few GeV 
than expected from other known sources \cite{TheFermi-LAT:2017vmf}. 
The analyses
of the GC gamma ray excess after considering different uncertainties puts an upper limit on the dark matter annihilation
cross section in terms of the DM mass and the annihilation channel. 

\begin{figure}
\centering
\includegraphics[scale=.35, angle=270]{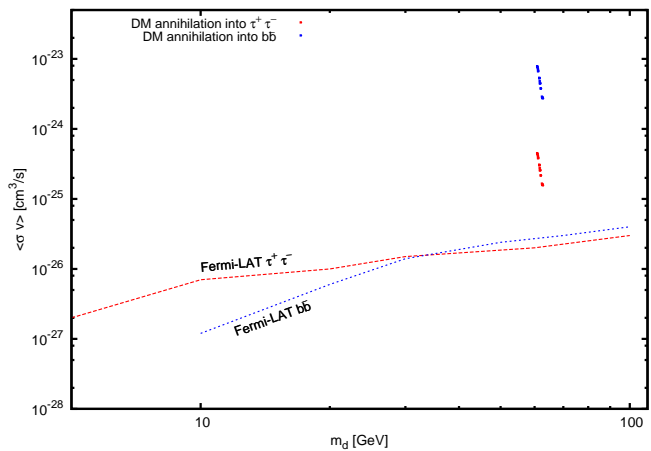}
\caption{The dark matter velocity-averaged annihilation cross section into $b\bar b$ (blue) and $\tau^+ \tau^-$ (red) for 
the DM mass up to $100$ GeV. For both channels the DM masses $61-62$ GeV shown in Fig. \ref{pseudo-plot} 
are excluded by the {\it Fermi}-LAT 
$b\bar b$ and $\tau^+ \tau^-$ upper limits.}
\label{indir}
\end{figure}

Here we examine our model  against the {\it Fermi}-LAT bounds on DM annihilation cross section 
for two representative channels $b\bar b$ and $\tau^+\tau^-$ which 
is relevant for DM masses up to $100$ GeV \cite{TheFermi-LAT:2017vmf}
\footnote{Many thanks to Andrea Albert and Dmitry Malyshev for providing me with the {\it Fermi}-LAT data on Fig. 34 
in \cite{TheFermi-LAT:2017vmf}}. 
For the current model we have computed the velocity-averaged annihilation 
cross section into two channels $b\bar b$ and $\tau^+ \tau^-$ for DM mass up to $100$ GeV. 
All other constraints 
considered in the last sections are imposed in this computation. It is evident from Fig. \ref{indir} 
that for both channels the DM
mass of $\sim 61-62$ GeV obtained before for the pseudoscalar mediator case in $T_c^+$ solution
(right plot in Fig. \ref{pseudo-plot}) is excluded by the {\it Fermi}-LAT limit. 
We have exploited the {\it Fermi}-LAT upper limit obtained from the generalized NFW dark matter density profile 
($\gamma=1.25$) in the center of the Galaxy.

\begin{figure}
\begin{minipage}[b]{.46\textwidth}
 \includegraphics[scale=.35, angle=270]{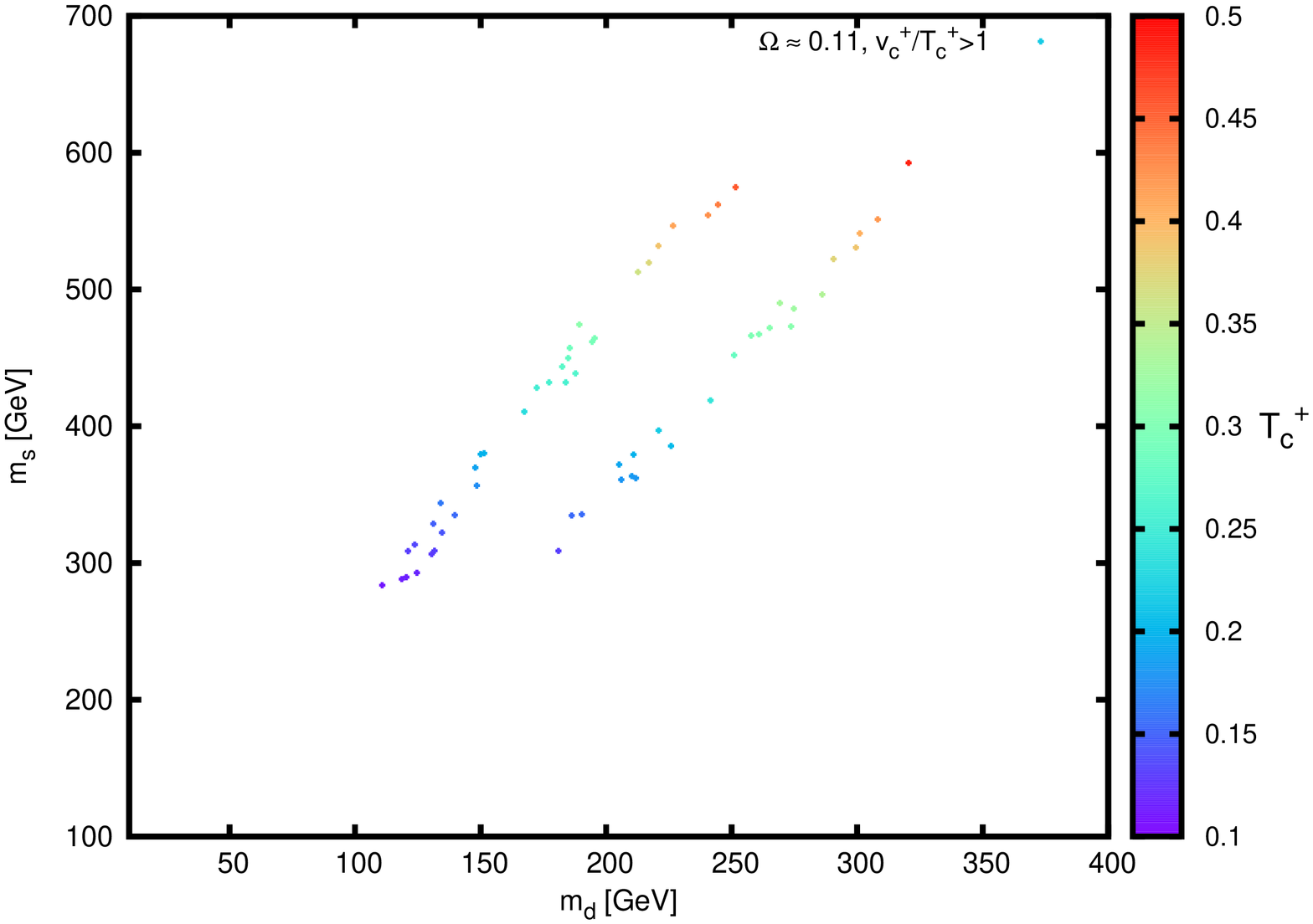}
\end{minipage}\hspace{1.5cm}
\begin{minipage}[b]{.46\textwidth}
 \includegraphics[scale=.35, angle=270]{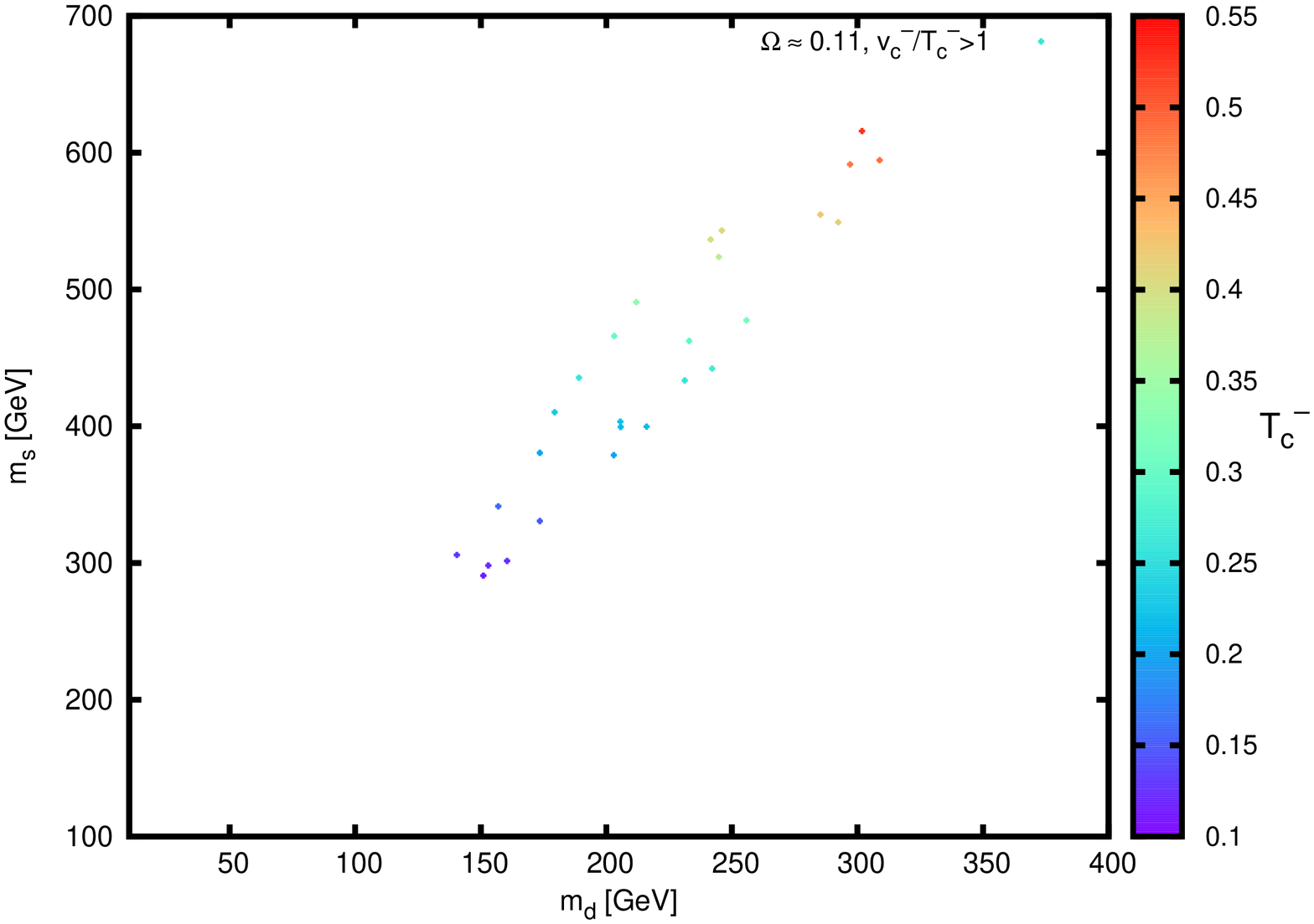}
\end{minipage}
\caption{The pseudoscalar mediator mass against the allowed DM mass for 
{\it left)} $T^+_c$ solution {\it right)} $T^-_c$ solution. The critical temperature is shown in the color spectrum.}
\label{mediator-DM} 
\end{figure}

After considering the {\it Fermi}-LAT constraint the viable DM mass with 
pseudoscalar mediator will be in the range $\sim 110-320$ GeV (see Fig. \ref{pseudo-plot}). 
Therefore the SM Higgs particle cannot decay into dark matter particle, hence the model is not bounded by
invisible Higgs decay constraint. In Fig. \ref{mediator-DM} the DM mass is depicted against the pseudoscalar mediator
mass considering all the aforementioned constraints. The points observed in this figure are as the following. First, 
for both $T_c^+$ and $T_c^-$ solutions the strongly first-order phase transition occurs in higher temperature 
for heavier dark matter particle. Second, the mass of the pseudoscalar mediator is in the range $290-620$ GeV. Again
the LHC exotic Higgs decay constraint which requires $m_s<m_h/2$ is not applicable here.

The mono-jet searches at the LHC could also restrict DM models specially for low DM mass and heavier mediators 
where the dark matter production cross section becomes larger. It is shown in \cite{Baek:2017vzd} that 
even for $m_s>2m_d$ the LHC mono-jet search limit \cite{Aaboud:2016tnv} 
does not constrain more the parameter space for the current model.

\section{Conclusion}
In this paper we have examined a fermionic dark matter model possessing a pseudoscalar or a scalar as the mediator
whether it could
explain the electroweak baryogenesis as well as evading the dark matter constraints. The (pseudo)scalar in the model is 
interpreted as a second Higgs-like particle beside the SM Higgs. We assumed that the scalars have zero 
vacuum expectation values in the symmetric phase but as the temperature comes down with the expansion of the
universe, the Higgs and the (pseudo)scalar undergo a non-zero vacuum expectation value and the electroweak phase transition 
takes place. We have obtained the critical temperature and the washout criterion analytically. Then we have 
computed the relic density using the {\tt micrOMEGAs} package while we have considered all the constraints for 
the electroweak phase transition to be first-order. The numerical results show that there exist a viable space of parameters
when the mediator is chosen to be a pseudoscalar with the dark matter mass in the range of $110-320$ GeV plus the resonance 
area with $m_d=61-62$ GeV. If 
the SM-DM mediator in the theory is taken to be a scalar we have shown that there is only a small region 
that can satisfy all the constraints including the direct detection bound from XENON100/LUX and that will be excluded with 
the new bounds from the experiments such as XENON1T. 

We then constrained more the model with the pseudoscalar mediator by the {\it Fermi}-LAT upper limit on DM annihilation into 
$b\bar b$ and $\tau^+ \tau^-$ which put a bound for DM mass up to $100$ GeV. We have shown that this bound excludes the 
DM mass $61-62$ GeV we obtained before and the DM mass in the viable space becomes 
$m_d=110-320$ GeV. The pseudoscalar mediator mass afterwards turned out to be $m_s=290-620$ GeV, that means the model 
cannot be constrained more from exotic and invisible Higgs decay bounds at the LHC. It has also 
been pointed out that the LHC mono-jet search limit does not affect the viable space of parameters.

\appendix
\section{(Pseudo)scalar One-loop Mass Correction by Dark Matter Fermion}\label{1loop}

The first three terms in $\kappa_{h}$ in eq. (\ref{kaph}) are the one-loop thermal correction
in the SM sector, see \cite{Carrington:1991hz}. The other terms in
$\kappa_{h}$ and all the coefficients in $\kappa_{s}$ stem from the singlet
(pseudo)scalar extension. For the details of the one-loop thermal calculations one may refer to {\cite{Chowdhury:2014tpa}. 
The coefficients $\lambda$ and $\lambda_s$ in eq. (\ref{kaps}) comes from the (pseudo)scalar mass correction with the Higgs or
the (pseudo)scalar 
respectively in the loop. The signs of these two coefficients do not change for the scalar or the pseudoscalar. 
The third coefficient in eq. (\ref{kaps}) is due to the (pseudo)scalar mass correction with the fermion 
dark matter in the loop. The sign of this term is however different 
for the scalar or the pseudoscalar. It will be plus for the scalar
and minus for the pseudoscalar. We elaborate this point here.

The contribution from the fermion (dark matter) loop for the pseudoscalar mass correction is the following,  
\begin{equation}\label{dm-}
 \delta m^{2} _{(F-)}=
g_{d}^{2}\int\frac{d^{4}k}{\left(2\pi\right)^{4}}\frac{\text{Tr}\left[\gamma^{5}\left(\gamma^{\mu}k_{\mu}
+\gamma^{\mu}p_{\mu}+m_{d}\right)\gamma^{5}\left(\gamma^{\nu}k_{\nu}
+m_{d}\right)\right]}{\left(\left(k+p\right)^{2}-m_{d}^{2}\right)\left(k^{2}-m_{d}^{2}\right)}\,,
\end{equation}
and the same correction for the scalar reads,  
\begin{equation}\label{dm+}
 \delta m^{2} _{(F+)}=
g_{d}^{2}\int\frac{d^{4}k}{\left(2\pi\right)^{4}}\frac{\text{Tr}\left[\left(\gamma^{\mu}k_{\mu}
+\gamma^{\mu}p_{\mu}+m_{d}\right)\left(\gamma^{\nu}k_{\nu}
+m_{d}\right)\right]}{\left(\left(k+p\right)^{2}-m_{d}^{2}\right)\left(k^{2}-m_{d}^{2}\right)}\,.
\end{equation}
Eqs. (\ref{dm-}) and (\ref{dm+}) can be written as,
\begin{equation}
\delta m^{2} _{(F\pm)}=g_{d}^{2}\int\frac{d^{4}k}{\left(2\pi\right)^{4}}\frac{\pm 4k^2\pm 4p.k+4m_d^2}
{\left(\left(k+p\right)^{2}-m_{d}^{2}\right)\left(k^{2}-m_{d}^{2}\right)}\,,
\end{equation}
where we have used the properties of the Dirac gamma matrices. The sign $+$ stands for the scalar and the sign $-$ 
stands for the pseudoscalar mass correction. At zero external momentum i.e. $p=0$ we have,  
\begin{equation}\label{p=0}
\delta m^{2} _{(F\pm)}=g_{d}^{2}
\int\frac{d^{4}k}{\left(2\pi\right)^{4}}\frac{\pm4k^2+4m_d^2}{\left(k^{2} -m_{d}^{2}\right)^2}\,.
\end{equation}
Now in thermal field theory terminology the following substitutions provide us with the one-loop correction at 
finite temperature (see e.g. \cite{das1997finite}),

\begin{equation}\label{tft}
\begin{split}
& k_0\rightarrow 2n i \pi T \,\,\,\,\,\,\,\,\,\,  \text{bosons}\\
& k_0\rightarrow (2n+1) i\pi T \,\,\,\,\,\,\,\,\,\, \text{fermions}\\
& \int{\frac{d^4k}{(2\pi)^4}} \rightarrow T \sum_n \int{\frac{d^3k}{(2\pi)^3}}\,.
\end{split}
\end{equation}

Using eqs. (\ref{p=0}) and (\ref{tft}) we get, 

\begin{equation}\label{thermal1}
\delta m^{2} _{(F+)}=-4 T g_{d}^{2} \sum_{n=\text{odd}} 
\int\frac{d^{3}k}{\left(2\pi\right)^{3}}\frac{1}{ n^2\pi^2 T^2
+{w}_k^2}-\frac{2m_d^2}{\left( n^2\pi^2 T^2
+{w}_k^2 \right)^2}\,,
\end{equation}
\begin{equation}\label{thermal2}
\delta m^{2} _{(F-)}=+4 T g_{d}^{2}\sum_{n=\text{odd}} 
\int\frac{d^{3}k}{\left(2\pi\right)^{3}}\frac{1}{ n^2\pi^2 T^2
+{w}_k^2} \,,
\end{equation}
where use has been made of ${w}_k^2=\overrightarrow{k}^2+m_d^2$. 
We see that the mass correction for the scalar 
contains a logarithmic divergence while for the pseudoscalar such a term 
does not exist. 

Making the summations in eqs. (\ref{thermal1}) and (\ref{thermal2}) by 
\begin{equation}
 \sum_{n=-\infty}^{+\infty}\frac{1}{n^2+x^2}=\frac{\pi}{x}\coth{\pi x} \,,
\end{equation}
ignoring the second term in eq. (\ref{thermal1}) which 
does not contribute in $\mathcal O (T^2)$ we arrive at, 
\begin{equation}\label{sumed}
 \delta m^{2} _{(F\pm)}=\mp 2 g_{d}^{2}
\int\frac{d^{3}k}{\left(2\pi\right)^{3}}\frac{1}{ 
{w}_k} \pm 4T g_{d}^{2}
\int\frac{d^{3}k}{\left(2\pi\right)^{3}} \frac{1}{ 
{w}_k} \frac{1}{ 1+e^{w_k/T}}\,.
\end{equation}
The first term in eq. (\ref{sumed}) is the zero temperature contribution and the second term is the correction at finite 
temperature. At high temperature approximation, $m/T\ll1$, this second term turns to be, 

\begin{equation}
 \delta m^{2}_{(F\pm)}(T)\approx \pm \frac{1}{6} g_d^2 T^2\,, 
\end{equation}
 which accounts for the minus sign in eq. (\ref{kaps}) for the pseudoscalar case and the plus sign for the scalar. 
 
\section*{Acknowledgment}
I would like to thank Karim Ghorbani, Alexander Pukhov and Nicolao Fornengo for very useful discussions. 
I am grateful for the hospitality and INFN-Torino university support when visiting the physics department of 
Torino university in 
February 2017 where this work was partially done. 
\bibliography{Ref}{}
\bibliographystyle{JHEP}
\end{document}